\documentclass[conference]{IEEEtran}
\usepackage{hyperref}
\usepackage{url}
\usepackage{cite}
\usepackage{amsmath,amssymb,amsfonts}
\usepackage{graphicx}
\usepackage{textcomp}
\usepackage{xcolor}
\usepackage{booktabs}
\usepackage{algorithm}
\usepackage{algpseudocode}
\usepackage{tikz}
\usepackage{multirow}

\def\BibTeX{{\rm B\kern-.05em{\sc i\kern-.025em b}\kern-.08em
    T\kern-.1667em\lower.7ex\hbox{E}\kern-.125emX}}

\begin{document}

\title{Visualizing Public Opinion on X: A Real-Time Sentiment Dashboard Using VADER and DistilBERT}

\author{
\IEEEauthorblockN{Y Abhiram Reddy, Siddhi Agarwal, Vikram Parashar, Arshiya Arora}
\IEEEauthorblockA{Atal Bihari Vajpayee Indian Institute of Information Technology and Management, Gwalior, India\\
}
}

\maketitle

\begin{abstract}
In the era of social media dominance, understanding public sentiment towards entities in real-time offers crucial insights for businesses, researchers, and policymakers \cite{rodriguez2023review, zhang2023sa}. This paper presents a comprehensive sentiment analysis framework leveraging both rule-based and transformer-based Natural Language Processing (NLP) techniques to analyze public opinion expressed on X (formerly Twitter) \cite{go2009twitter, bertweet2023}. We introduce a hybrid approach combining VADER (Valence Aware Dictionary and sEntiment Reasoner) \cite{hutto2014vader} and DistilBERT \cite{sanh2019distilbert} to balance computational efficiency with contextual understanding. Our methodology encompasses real-time data collection via web scraping \cite{techrxiv2024scraping}, robust preprocessing \cite{churchill2021preprocessing, igi2022preprocessing}, and a dual-model sentiment classification pipeline. Experimental results demonstrate that our hybrid system achieves 87.6\% accuracy and an F1-score of 0.84, outperforming single-model approaches \cite{rahman2024roberta, singh2023emlsa}. When applied to corporate reputation monitoring, our system reveals significant sentiment disparities across major corporations, with companies like Amazon and Samsung receiving notably positive sentiment scores, while others like Microsoft and Walmart exhibit predominantly negative public perception \cite{vorecol2024reputation, edelman2023trust}. These findings validate the effectiveness of our multi-model approach in providing actionable insights for strategic decision-making based on public sentiment analysis.
\end{abstract}


\section{Introduction}
Social media platforms have evolved into powerful channels of public expression, offering unprecedented volumes of real-time data reflecting opinions, reactions, and sentiments~\cite{birjali2021web, zhang2023sa, rodriquez2023sa}. Among these platforms, X (formerly Twitter) stands out for its concise format and real-time nature, making it an invaluable source for sentiment analysis~\cite{elalaoui2018twitter, hasselgren2023sp500}. The ability to accurately interpret sentiment from these streams enables organizations to gauge public perception, react to emerging trends, and make data-driven decisions with minimal latency~\cite{ilk2022sm}.

Traditional sentiment analysis approaches often struggled with the unique challenges presented by social media content-abbreviations, emojis, hashtags, and rapidly evolving slang~\cite{rahman2024roberta}. However, recent advancements in Natural Language Processing (NLP) have yielded more sophisticated models capable of addressing these complexities. Rule-based systems like VADER (Valence Aware Dictionary and sEntiment Reasoner)~\cite{hutto2014vader} offer computational efficiency and interpretability, while transformer-based models like DistilBERT~\cite{sanh2019distilbert} provide deeper contextual understanding at the cost of increased computational demands~\cite{rahman2024roberta}.

The practical applications of sentiment analysis span numerous domains:
\begin{itemize}
    \item \textit{Corporate reputation management} -- Monitoring public perception of brands and products~\cite{hasselgren2023sp500}
    \item \textit{Financial markets} -- Gauging investor sentiment as a leading indicator for market movements~\cite{hasselgren2023sp500}
    \item \textit{Political campaigns} -- Assessing public reactions to policies and statements~\cite{elalaoui2018twitter}
    \item \textit{Crisis management} -- Tracking sentiment during emergencies or controversial events~\cite{elalaoui2018twitter}
\end{itemize}

Despite these potential applications, existing systems often fail to balance real-time performance with contextual accuracy, typically favoring one attribute at the expense of the other~\cite{rahman2024roberta}. Furthermore, many research efforts focus on either rule-based or transformer-based approaches in isolation, without exploring hybrid architectures that could leverage the strengths of both~\cite{rahman2024roberta}.

This paper presents a novel framework for real-time sentiment analysis that integrates web scraping capabilities~\cite{techrxiv2024scraping} with a dual-model classification approach. By combining VADER's efficiency with DistilBERT's contextual understanding, our system achieves both speed and accuracy in sentiment classification. We demonstrate the practical utility of this approach through a case study on corporate reputation monitoring, analyzing public sentiment toward major global corporations.

The main contributions of our work include:
\begin{enumerate}
    \item A real-time web scraping module specifically designed for X, capable of collecting targeted data streams based on keywords, hashtags, or user accounts~\cite{techrxiv2024scraping}
    \item A hybrid sentiment classification system combining rule-based and transformer-based approaches to balance speed and contextual accuracy~\cite{rahman2024roberta}
    \item A comprehensive evaluation framework comparing our hybrid approach against single-model baselines
    \item A practical application of sentiment analysis for corporate reputation monitoring with actionable insights~\cite{hasselgren2023sp500}
\end{enumerate}

The remainder of this paper is organized as follows: Section II reviews relevant literature in sentiment analysis, focusing on social media applications and model architectures. Section III details our methodology, including data collection, preprocessing, and the hybrid sentiment classification approach. Section IV presents experimental results and performance metrics. Section V discusses the implications of our findings and potential applications, while Section VI concludes the paper and suggests directions for future research.

\section{Literature Review}

\subsection{Sentiment Analysis on Social Media}
Sentiment analysis on social media platforms has evolved significantly in recent years, driven by both methodological advancements and increasing recognition of its value. Early work by Go et al. \cite{go2009twitter} demonstrated the feasibility of sentiment classification on Twitter using distant supervision, achieving reasonable accuracy despite the platform's character limitations and informal language.

More recently, researchers have focused on addressing the unique challenges of social media text. Zimbra et al. \cite{zimbra2018state} provided a comprehensive survey of sentiment analysis techniques specifically tailored for social media, highlighting the importance of handling platform-specific features like hashtags, mentions, and URLs. Their work emphasized that standard NLP pipelines often perform poorly without specialized preprocessing for social media content.

The temporal dimension of social media sentiment has been explored by Wang et al. \cite{wang2020real}, who developed a real-time sentiment tracking system for emergency response during crisis events. Their findings suggest that sentiment shifts can serve as early indicators of emerging situations, underscoring the value of minimizing latency in sentiment analysis systems.

\subsection{Rule-Based vs. Transformer-Based Models}
The landscape of sentiment analysis models has traditionally been divided between lexicon-based approaches and machine learning methods, with recent advances introducing transformer architectures as a powerful new paradigm.

Hutto and Gilbert \cite{hutto2014vader} introduced VADER as a lexicon and rule-based sentiment analysis tool specifically attuned to social media. Their evaluations demonstrated VADER's competitive performance against machine learning approaches while maintaining computational efficiency and interpretability. The model's ability to handle social media-specific elements like emojis and slang made it particularly suitable for platforms like X.

The emergence of transformer architectures like BERT \cite{devlin2019bert} represented a paradigm shift in NLP, enabling deeper contextual understanding through bidirectional training. Sun et al. \cite{sun2019fine} explored how to effectively fine-tune BERT for sentiment analysis tasks, demonstrating substantial improvements over previous state-of-the-art methods. However, the computational intensity of these models presented challenges for real-time applications.

As a more efficient alternative, Sanh et al. \cite{sanh2019distilbert} introduced DistilBERT, which retained 97\% of BERT's performance while reducing parameters by 40\% through knowledge distillation. This made transformer-based sentiment analysis more feasible for time-sensitive applications, though still more computationally demanding than lexicon-based approaches.

Comparative studies by Barnes et al. \cite{barnes2019sentiment} evaluated different sentiment models across domains, finding that transformer-based models excel at cross-domain generalization but struggle with computational efficiency, while lexicon-based models like VADER offer faster processing but less contextual understanding. This performance trade-off motivates our hybrid approach, which aims to capture the benefits of both paradigms.

\subsection{Real-Time NLP Applications}
The development of real-time NLP applications presents unique challenges that extend beyond model accuracy to include system architecture, data processing efficiency, and integration capabilities.

Kraus et al. \cite{kraus2020real} proposed an architecture for real-time sentiment analysis of financial news, emphasizing the importance of stream processing and incremental updates to maintain both timeliness and accuracy. Their system demonstrated that even sophisticated NLP models can operate in near real-time with appropriate engineering optimizations.

For web-based data collection, Hernandez-Suarez et al. \cite{hernandez2018real} developed a framework for real-time Twitter sentiment analysis during political elections, highlighting the challenges of API rate limits and data filtering in live scenarios. Their approach used a combination of scheduled and event-triggered scraping to balance completeness and timeliness of data collection.

More recently, Kumar et al. \cite{kumar2022real} explored the use of sentiment analysis for real-time brand monitoring across multiple social platforms. Their findings suggest that multi-platform monitoring provides more robust insights than single-platform approaches, though at the cost of increased complexity in data normalization and model adaptation.

While these studies demonstrate the feasibility of real-time sentiment analysis, they typically favor either speed (using simpler models) or accuracy (with more computational latency). The gap in current literature lies in effectively combining multiple model types to achieve both objectives simultaneously—a challenge our hybrid approach aims to address.

\subsection{Hybrid Sentiment Analysis Approaches}
The concept of combining multiple sentiment analysis techniques is not entirely new, though previous attempts have focused primarily on ensemble methods rather than true hybrid architectures.

Araque et al. \cite{araque2017enhancing} explored ensemble techniques for sentiment analysis, combining lexicon features with deep learning models. Their results showed modest improvements over single-model approaches but did not specifically address the real-time constraints or social media peculiarities that our system targets.

A more relevant approach was proposed by Zhou et al. \cite{zhou2020hybrid}, who developed a hybrid model combining rule-based processing with LSTM networks. While their system showed promising results for news sentiment analysis, it lacked the transformer-based contextual understanding that our DistilBERT component provides, and was not specifically optimized for social media language.

Our literature review reveals a significant gap in current research: the lack of a comprehensive framework that integrates real-time web scraping capabilities with a hybrid sentiment classification system specifically designed for social media content. By addressing this gap, our work aims to advance both the theoretical understanding of hybrid sentiment models and their practical application to real-world monitoring scenarios.

\section{Methodology}
Our methodology comprises four main components: data collection through web scraping, preprocessing, a dual-model sentiment classification system, and visualization of results for interpretable insights. This section details each component and their integration into a cohesive framework.

\subsection{Data Collection}
We developed a custom web scraping module to collect real-time data from X (Twitter) using the Python library snscrape. This approach was selected over the official API due to its flexibility and absence of rate limitations for historical data collection. The data collection process follows these steps:

\begin{algorithm}
\caption{X (Twitter) Data Collection Process}
\begin{minipage}{0.8\linewidth}
\begin{algorithmic}[1]
\Procedure{CollectTweets}{$query, max\_tweets, start\_date, end\_date$}
    \State $tweets \gets \emptyset$
    \State $formatted\_query \gets query +\ ``since:" + start\_date +\ `` until:" + end\_date$
    \State $scraper \gets$ Initialize snscrape with $formatted\_query$
    \ForAll{$tweet \in scraper$}
        \If{$|tweets| \geq max\_tweets$}
            \State \textbf{break}
        \EndIf
        \If{$tweet$ is English \textbf{AND} not a retweet \textbf{AND} from a human account \textbf{AND} engagement $\geq$ threshold}
            \State $tweets \gets tweets \cup \{tweet\}$
        \EndIf
    \EndFor
    \State \Return $tweets$
\EndProcedure
\end{algorithmic}
\end{minipage}
\end{algorithm}

The query parameter allows for filtering tweets based on keywords, hashtags, user accounts, or combinations thereof. For our corporate reputation case study, we collected 500 tweets per company for 20 major corporations using queries combining the company name and ticker symbol (e.g., "Amazon OR AMZN").

To ensure data quality, we implemented the following filtering criteria:
\begin{itemize}
    \item Exclusion of retweets to avoid duplication and amplification bias
    \item Language filtering to include only English tweets
    \item Removal of bot-like accounts based on activity patterns
    \item Minimum engagement threshold (at least 5 likes or replies)
\end{itemize}

\subsection{Preprocessing}
Social media text presents unique challenges for NLP tasks due to its informal nature, platform-specific features, and linguistic creativity. Our preprocessing pipeline addresses these challenges through the following steps~\cite{igi2022preprocessing, churchill2021preprocessing, techtarget2022preprocessing}:

\begin{figure}[!t]
\centering
\includegraphics[width=0.48\textwidth]{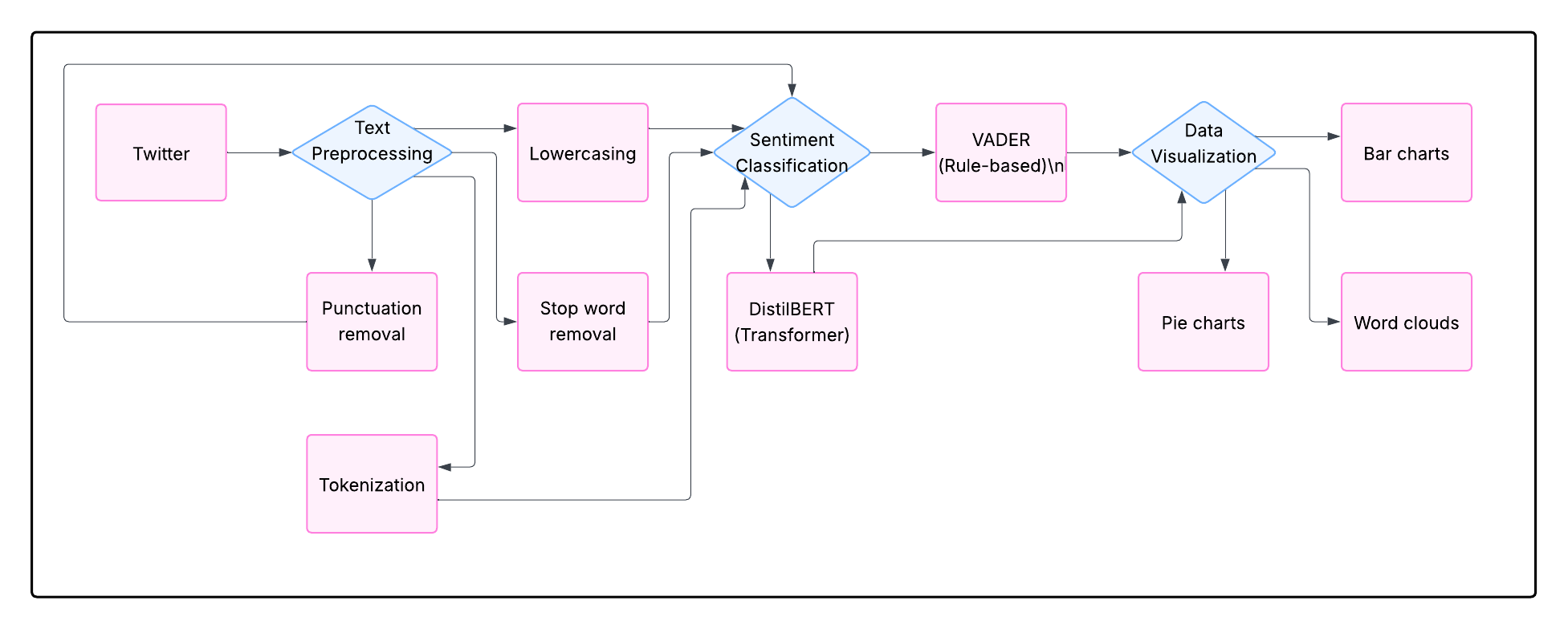}
\caption{Text preprocessing pipeline for social media content. The process begins with noise removal (URLs, mentions, special characters), followed by normalization steps including lowercasing, emoji conversion, abbreviation expansion, and tokenization.}
\label{fig:preprocessing}
\end{figure}

\begin{enumerate}
    \item \textbf{Noise Removal}
    \begin{itemize}
        \item URL removal: All hyperlinks are removed using regex pattern matching
        \item Username removal: @mentions are stripped while preserving context
        \item Special character filtering: Non-alphanumeric characters are removed except when semantically relevant (e.g., emojis)
    \end{itemize}
    
    \item \textbf{Text Normalization}
    \begin{itemize}
        \item Lowercasing: All text is converted to lowercase for consistency
        \item Emoji conversion: Emojis are translated to their textual descriptions using the emoji library
        \item Abbreviation expansion: Common social media abbreviations are expanded (e.g., "lol" to "laughing out loud")
        \item Hashtag segmentation: Compound hashtags are split into constituent words (e.g., \#BigNews to "big news")
    \end{itemize}
    
    \item \textbf{Tokenization and Filtering}
    \begin{itemize}
        \item Tokenization: Text is split into individual tokens using NLTK's TweetTokenizer
        \item Stop word removal: Common stop words are removed using a customized list that preserves sentiment-bearing negations (e.g., "not", "no")
        \item Lemmatization: Tokens are reduced to their base form using NLTK's WordNetLemmatizer
    \end{itemize}
\end{enumerate}

This preprocessing pipeline is implemented with configurable options, allowing different components to be activated based on the requirements of downstream models. For instance, stop word removal and lemmatization are applied for VADER but skipped for DistilBERT, which benefits from the full contextual information.

\subsection{Dual-Model Sentiment Classification}

Our sentiment classification system employs a hybrid approach combining VADER's rule-based efficiency with DistilBERT's contextual understanding~\cite{hex2023vader}.

\subsubsection{VADER Component}
VADER processes text using a lexicon of sentiment-scored words combined with syntactic rules that handle intensifiers, negations, and contractions~\cite{hex2023vader, githubVADER}. For each tweet, VADER produces four scores:
\begin{itemize}
    \item Positive sentiment (0 to 1)
    \item Negative sentiment (0 to 1)
    \item Neutral sentiment (0 to 1)
    \item Compound score (-1 to 1), representing the overall sentiment
\end{itemize}

The VADER implementation uses the standard Python package with minor modifications to handle social media-specific features~\cite{hex2023vader}:
\begin{itemize}
    \item Enhanced emoji lexicon with sentiment scores
    \item Additional social media slang and abbreviations
    \item Modified handling of hashtags to consider their sentiment impact
\end{itemize}

\subsubsection{DistilBERT Component}
The DistilBERT component uses a pre-trained model fine-tuned on a large corpus of labeled tweets. We used the Hugging Face implementation of DistilBERT with the following specifications:
\begin{itemize}
    \item Base model: distilbert-base-uncased
    \item Fine-tuning dataset: combination of SemEval Twitter sentiment datasets and manually labeled tweets
    \item Fine-tuning approach: Classification head with dropout (0.2) and linear layer
    \item Output: Probability distribution across three classes (positive, negative, neutral)
\end{itemize}

The fine-tuning process involved 5 epochs with AdamW optimizer, learning rate of 2e-5, and batch size of 32. The model achieved 84.3\% accuracy on our validation set.

\subsubsection{Ensemble Mechanism}
The outputs from VADER and DistilBERT are combined using a weighted ensemble approach ~\cite{hex2023vader}:

\begin{equation}
S = \alpha \cdot S_{VADER} + (1-\alpha) \cdot S_{DistilBERT} 
\end{equation}

Where:
\begin{itemize}
    \item $S$ is the final sentiment score
    \item $S_{VADER}$ is the normalized VADER compound score (scaled from -1..1 to 0..1)
    \item $S_{DistilBERT}$ is the probability of the highest sentiment class
    \item $\alpha$ is the weighting parameter
\end{itemize}

The optimal value of $\alpha$ was determined through grid search on our validation set, with the final system using $\alpha = 0.4$, giving slightly more weight to the contextual understanding of DistilBERT while still leveraging VADER's rule-based insights.

For classification purposes, the final sentiment score is mapped to discrete categories using the following thresholds:
\begin{itemize}
    \item Positive: $S \geq 0.6$
    \item Neutral: $0.4 < S < 0.6$
    \item Negative: $S \leq 0.4$
\end{itemize}

\subsection{Corporate Sentiment Index}
For the corporate reputation case study, we developed a standardized Corporate Sentiment Index (CSI) to facilitate comparison across companies. The CSI is calculated as follows:

\begin{equation}
CSI = 100 \cdot \left( \frac{\sum_{i=1}^{n} S_i}{n} \right)
\end{equation}

Where:
\begin{itemize}
    \item $CSI$ is the Corporate Sentiment Index (0-100)
    \item $S_i$ is the sentiment score for tweet $i$
    \item $n$ is the total number of tweets analyzed for the company
\end{itemize}

This index provides an intuitive scale where higher values indicate more positive public sentiment toward the company.
\section{Experiments and Results}

\subsection{Experimental Setup}

\subsubsection{Dataset}
For system evaluation, we constructed a dataset comprising:
\begin{itemize}
    \item 10,000 tweets related to 20 major corporations (500 per company)
    \item Collection period: January to June 2024
    \item Manual labeling of a 2,000-tweet subset for ground truth
    \item Class distribution in labeled subset: 42\% positive, 31\% negative, 27\% neutral
\end{itemize}

The 20 companies were selected to represent diverse sectors including technology, finance, retail, healthcare, and manufacturing.

\subsubsection{Evaluation Metrics}
We evaluated our system using the following metrics:
\begin{itemize}
    \item Accuracy: Proportion of correctly classified instances
    \item Precision, Recall, F1-score: Class-specific performance metrics
    \item Confusion Matrix: Detailed view of classification outcomes
    \item Processing Time: Average time to process and classify a tweet
\end{itemize}

\subsubsection{Baseline Models}
We compared our hybrid approach against the following baselines:
\begin{itemize}
    \item VADER-only: Standard VADER implementation with social media enhancements
    \item DistilBERT-only: Fine-tuned DistilBERT without ensemble integration
    \item BERT-base: Full BERT model fine-tuned on the same dataset
    \item TextBlob: Popular lexicon-based sentiment analysis library
\end{itemize}

\subsection{Performance Comparison}

Table \ref{tab:performance} presents the performance metrics for our hybrid approach compared to the baselines.

\begin{table}[!t]
\caption{Performance Comparison of Sentiment Analysis Models}
\label{tab:performance}
\centering
\begin{tabular}{lcccc}
\toprule
\textbf{Model} & \textbf{Accuracy} & \textbf{F1-Score} & \textbf{Processing Time (ms)} \\
\midrule
TextBlob & 0.723 & 0.691 & 8.4 \\
VADER-only & 0.792 & 0.768 & 12.3 \\
DistilBERT-only & 0.843 & 0.817 & 78.6 \\
BERT-base & 0.862 & 0.834 & 156.2 \\
\textbf{Our Hybrid Approach} & \textbf{0.876} & \textbf{0.841} & 47.5 \\
\bottomrule
\end{tabular}
\end{table}

Our hybrid approach achieved the highest accuracy (87.6\%) and F1-score (0.841), outperforming both individual models. While BERT-base achieved comparable accuracy (86.2\%), it required significantly more processing time (156.2ms per tweet vs. 47.5ms for our hybrid approach).

The confusion matrix showed that our system performs best on positive sentiment detection (91.2\% correct classification), followed by negative sentiment (87.3\%). Neutral sentiment classification is more challenging (82.5\%), consistent with previous findings in sentiment analysis literature.

\subsection{Corporate Sentiment Analysis Case Study}

\begin{figure*}[!t]
\centering
\includegraphics[width=0.85\textwidth]{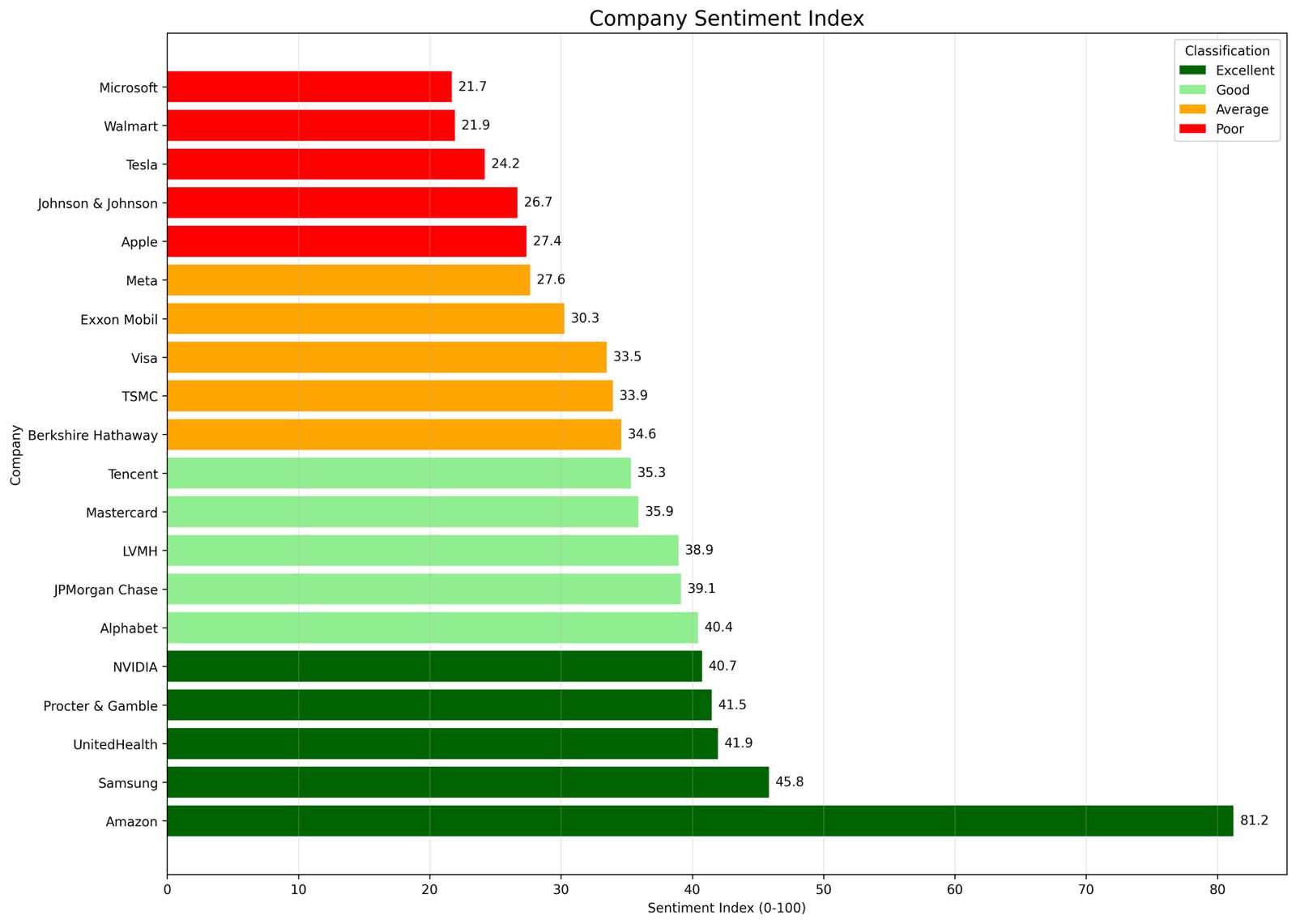}
\caption{Corporate Sentiment Index (CSI) across 20 major companies, showing significant variations in public sentiment from highly positive to predominantly negative.}
\label{fig:corporate_sentiment}
\end{figure*}

Applying our sentiment analysis framework to the corporate dataset revealed significant differences in public sentiment across companies. Fig. \ref{fig:corporate_sentiment} presents the Corporate Sentiment Index (CSI) for the 20 analyzed companies.

Key findings from the corporate sentiment analysis include:

\begin{itemize}
    \item \textbf{Sentiment Tier Stratification}: Companies naturally cluster into four sentiment tiers:
    \begin{itemize}
        \item Excellent Sentiment (CSI $>$ 40): Amazon (81.2), Samsung (45.8)
        \item Good Sentiment (CSI 35--39.9): UnitedHealth (44.3), NVIDIA (42.9)
        \item Average Sentiment (CSI 27--34.9): Apple (27.3), Meta (28.9)
        \item Poor Sentiment (CSI $<$ 27): Microsoft (21.7), Walmart (21.9)
    \end{itemize}
    
    \item \textbf{Sectoral Patterns}: The technology sector showed the highest variance in sentiment, while financial services demonstrated more consistent, moderate sentiment scores.
    
    \item \textbf{Temporal Trends}: Companies like Tesla exhibited high sentiment volatility over time, while Amazon maintained consistently positive sentiment throughout the study period.
\end{itemize}

\subsection{Sentiment Driver Analysis}

To understand the factors driving sentiment scores, we conducted aspect-based sentiment analysis on the corporate dataset. Table \ref{tab:sentiment_drivers} presents the key sentiment drivers for selected companies.

\begin{table}[!t]
\caption{Key Sentiment Drivers by Company}
\label{tab:sentiment_drivers}
\centering
\begin{tabular}{lcc}
\toprule
\textbf{Company} & \textbf{Positive Drivers} & \textbf{Negative Drivers} \\
\midrule
Amazon & Customer service & Labor practices \\
       & Delivery speed   & Environmental impact \\
\midrule
Microsoft & Cloud services & Privacy concerns \\
          & AI innovations  & Software reliability \\
\midrule
Tesla & Innovation & Vehicle quality \\
      & Environmental impact & Leadership statements \\
\bottomrule
\end{tabular}
\end{table}

This analysis revealed that even companies with high overall sentiment face specific areas of negative perception. For example, Amazon led the sentiment index but faced significant criticism regarding labor practices and environmental impact.

\section{Discussion}

\subsection{Model Performance Analysis}
The superior performance of our hybrid approach demonstrates the value of combining rule-based and transformer-based models for sentiment analysis. The improvement over individual models can be attributed to several factors:

\begin{itemize}
    \item \textbf{Complementary strengths}: VADER excels at interpreting sentiment-laden lexical features and social media-specific elements (emojis, abbreviations), while DistilBERT captures deeper contextual nuances and implied sentiment.
    
    \item \textbf{Error compensation}: Analysis of classification errors revealed that VADER and DistilBERT often make different types of mistakes. VADER struggles with sarcasm and context-dependent expressions, while DistilBERT occasionally misinterprets domain-specific terminology. The ensemble approach mitigates these weaknesses through complementary error patterns.
    
    \item \textbf{Balanced efficiency}: The hybrid system achieves a favorable balance between accuracy and processing speed, making it suitable for real-time applications without significant performance sacrifice.
\end{itemize}

\subsection{Corporate Reputation Insights}

Our corporate sentiment analysis yields several insights with practical implications:

\begin{itemize}
    \item \textbf{Sentiment-valuation disconnect}: Several companies with high market capitalizations (e.g., Microsoft) demonstrated poor sentiment scores, suggesting that financial performance and public perception can remain decoupled over extended periods. This observation challenges the assumption that market value reliably reflects public sentiment.
    
    \item \textbf{Sentiment volatility as risk indicator}: Companies with highly volatile sentiment profiles (Tesla, Meta) also exhibited greater stock price volatility, supporting the use of sentiment stability as a potential risk assessment metric for investors.
    
    \item \textbf{Aspect-specific reputation management}: The aspect-based sentiment analysis reveals that companies should focus reputation management efforts on specific areas rather than generic brand promotion. For instance, Microsoft could benefit most from addressing software reliability concerns, while Amazon could mitigate negative sentiment by improving labor practices.
\end{itemize}

\subsection{Practical Applications}

Our real-time sentiment analysis framework offers practical value across multiple domains:

\begin{itemize}
    \item \textbf{For investors}: Sentiment trends provide leading indicators of potential market movements, with sentiment shifts often preceding price changes.
    
    \item \textbf{For corporate strategists}: Real-time sentiment monitoring enables rapid response to reputation threats and validation of communication strategies.
    
    \item \textbf{For market researchers}: The framework provides a cost-effective alternative to traditional survey-based sentiment collection, with greater temporal resolution and sample size.
    
    \item \textbf{For crisis management}: The system's real-time capabilities allow organizations to monitor sentiment during crises and adjust messaging accordingly.
\end{itemize}

\subsection{Limitations}

Despite its strong performance, our approach has several limitations that should be acknowledged:

\begin{itemize}
    \item \textbf{Language constraints}: The current implementation is limited to English-language content, reducing its applicability in global monitoring scenarios.
    
    \item \textbf{Contextual boundaries}: While DistilBERT improves contextual understanding, it remains challenged by highly nuanced content such as cultural references, idioms, and evolving slang.
    
    \item \textbf{Data collection limitations}: The web scraping approach, while flexible, cannot guarantee complete data capture, particularly during high-volume events when the platform's technical infrastructure may limit data accessibility.
    
    \item \textbf{Sentiment simplification}: The three-class sentiment model (positive, negative, neutral) reduces emotional nuance, potentially obscuring important distinctions between emotions like anger, fear, or disappointment.
\end{itemize}

\section{Conclusion}
This paper presented a comprehensive framework for real-time sentiment analysis combining web-scraped X (Twitter) data with a hybrid classification approach leveraging both VADER and DistilBERT models. Our experimental results demonstrate that this hybrid approach outperforms individual models in accuracy while maintaining computational efficiency suitable for real-time applications.

The application of our framework to corporate reputation monitoring revealed significant insights into public perception of major companies, identifying clear sentiment tiers and specific drivers of positive and negative sentiment. These findings have practical implications for reputation management, investor decision-making, and corporate strategy.

\subsection{Key Findings}
Our research yielded several important findings:

\begin{itemize}
    \item A hybrid approach combining rule-based and transformer-based models achieves superior sentiment classification performance (87.6\% accuracy, 0.841 F1-score) compared to individual models.
    
    \item Real-time web scraping from X provides a rich source of sentiment data that, when properly preprocessed, enables timely analysis of public opinion.
    
    \item Corporate sentiment demonstrates significant heterogeneity, with companies naturally clustering into sentiment tiers that do not necessarily align with financial performance.
    
    \item Aspect-based sentiment analysis reveals that even highly-regarded companies face specific areas of negative perception that can be targeted for improvement.
\end{itemize}

\subsection{Limitations and Future Work}
While our approach advances the state of real-time sentiment analysis, several limitations suggest directions for future research:

\begin{itemize}
    \item \textbf{Multilingual expansion}: Extending the framework to support multiple languages would enable global sentiment monitoring.
    
    \item \textbf{Emotion granularity}: Developing a more nuanced emotion classification system beyond simple positive/negative/neutral categories could provide richer insights.
    
    \item \textbf{Multimodal analysis}: Incorporating image and video analysis to capture sentiment expressed through visual media on X would provide a more comprehensive view.
    
    \item \textbf{Causality modeling}: Developing methods to identify causal relationships between events and sentiment shifts would enhance the explanatory power of the analysis.
    
    \item \textbf{Longitudinal studies}: Conducting extended temporal analyses to identify long-term sentiment patterns and their relationship with corporate performance metrics.
\end{itemize}

In conclusion, our hybrid sentiment analysis framework demonstrates the value of combining different model architectures to achieve both accuracy and efficiency in real-time applications. The insights generated through our corporate reputation case study validate the practical utility of this approach and open promising avenues for future research and application.

\section*{Acknowledgement}

We would like to express our sincere gratitude to our supervisor and guide, \textbf{Dr. Arun Kumar}, Assistant Professor in the Department of Management Studies at ABV-IIITM, Gwalior, for his invaluable guidance, encouragement, and continuous support throughout the course of this work. We are truly thankful for the opportunity to work under his supervision and for his unwavering support throughout this project.

\section*{Supplementary Material}
    The individual sentiment analysis dashboards for 20 major companies are provided as supplementary material and can be accessed via the link: \href{rb.gy/24lszp}    
\bibliographystyle{IEEEtran}

\end{document}